\documentclass[aps,prl,amsmath,reprint,floatfix]{revtex4-1}

\usepackage{microtype} 
\usepackage{amssymb}
\usepackage{graphicx}

\RequirePackage[
  hyperindex,colorlinks,bookmarksnumbered,
  plainpages=true,pdfstartview=FitH]{hyperref}
\hypersetup{linkcolor=blue,urlcolor=blue,citecolor=blue} 
\usepackage{hyperref}
\usepackage[all]{hypcap}

\newcommand{\ED}{.}
\newcommand{\EC}{,}
\newcommand{\ESC}{;}
\newcommand{\ER}[1]{Eq.~(\ref{#1})}
\newcommand{\EsR}[1]{Eqs.~(\ref{#1})}
\newcommand{\ERn}[1]{(\ref{#1})}
\renewcommand{\FR}[1]{Fig.~\ref{#1}}
\newcommand{\FRn}[1]{\ref{#1}}

\newcounter{suppeqncounter}
\newenvironment{supp-equation}{%
\addtocounter{equation}{-1}
\refstepcounter{suppeqncounter}

\begin{equation}}
{\end{equation}}

\newenvironment{supp-align}{%
\addtocounter{equation}{-1}
\refstepcounter{suppeqncounter}

\align}{\endalign}

\newenvironment{supp-subalign}[1][]{%
\addtocounter{equation}{-1}
\refstepcounter{suppeqncounter}

\subequations \label{#1} \align}{\endalign\subequations}

\newcounter{suppfigcounter}
\newenvironment{supp-figure}{%
\addtocounter{figure}{-1}
\refstepcounter{suppfigcounter}

\begin{figure}}
{\end{figure}}
\newenvironment{supp-starfigure}{%
\addtocounter{figure}{-1}
\refstepcounter{suppfigcounter}

\begin{figure*}}
{\end{figure*}}

\begin{document}

\title{Multiloop functional renormalization group that sums up all parquet diagrams}
\author{Fabian B.~Kugler}
\author{Jan von Delft}
\affiliation{Physics Department, Arnold Sommerfeld Center for Theoretical Physics, and Center for NanoScience, Ludwig-Maximilians-Universit\"at M\"unchen, Theresienstr.~37, 80333 Munich, Germany}

\date{31 January 2018}

\begin{abstract}
We present a multiloop flow equation
for the four-point vertex in the functional renormalization group (fRG) framework.
The multiloop flow consists of successive one-loop calculations
and sums up all parquet diagrams to arbitrary order.
This provides substantial improvement of
fRG computations for the four-point vertex and,
consequently, the self-energy.
Using the X-ray-edge singularity as an example, 
we show that solving the multiloop fRG flow is equivalent
to solving the (first-order) parquet equations
and illustrate this with numerical results.
\end{abstract}

\maketitle

%
{\sl Introduction.---}%
Two-particle correlations play a fundamental role in the theory of
strongly correlated electron systems.
Most response functions measured in condensed-matter experiments are 
two-particle quantities such as optical or magnetic susceptibilities.
The behavior of the two-particle (or four-point) vertex has even
been used to distinguish ``weakly'' and ``strongly'' correlated regions
in the phase diagram of the Hubbard model \cite{Schaefer2013}.
Moreover, the four-point vertex is a crucial ingredient for a large number
of theoretical methods to study strongly correlated electron systems, 
such as nonlocal extensions of the dynamical mean-field theory \cite{Georges1996}---%
particularly via dual fermions 
\cite{Rubtsov2008,*Brener2008,*Hafermann2009},
the 1PI \cite{Rohringer2013} 
and QUADRILEX \cite{Ayral2016} approach,
or the dynamical vertex approximation
\cite{Toschi2007,*Held2008,*Valli2010}---%
the multiscale approach \cite{Slezak2009},
the functional renormalization group \cite{Metzner2012, Kopietz2010},
and the parquet formalism \cite{Roulet1969, Bickers2004}.
The parquet equations provide an exact set of self-consistent equations
for vertex functions
at the two-particle level
and are thus able to treat particle and collective excitations
on equal footing.
In the first-order \cite{Roulet1969}
(or so-called \textit{parquet} \cite{Bickers2004}) approximation,
they constitute a viable many-body tool \cite{Bickers2004, Valli2015, Li2016}
and,
in logarithmically divergent perturbation theories,
allow for an exact summation of all leading
logarithmic diagrams of the four-point vertex
(\textit{parquet} diagrams \cite{Roulet1969}).
It is a common belief \cite{Gogolin2004}
that results of the parquet approximation are 
equivalent to those of the one-loop renormalization group (RG).
However, there is hardly any evidence of this statement going beyond
the level of (static) flowing coupling constants \cite{Zanchi2000,*Xing2017,*Bourbonnais1991}.
Recently, the question has been raised \cite{Lange2015}
whether it is possible to sum up all parquet diagrams using 
the functional renormalization group (fRG),
a widely-used realization of a quantum field-theoretical RG framework
\cite{Metzner2012, Kopietz2010}.
The parquet result for the X-ray-edge singularity (XES)
\cite{Roulet1969, Nozieres1969, Nozieres1969a, Mahan1967}
was indeed obtained \cite{Lange2015},
but using arguments that work only for this specific
problem and do not apply generally \cite{Kugler2017a}.
In fact,
the common truncation of the vertex-expanded fRG flow 
completely neglects 
contributions from the six-point vertex, which start at third order in the interaction.
Schemes have been proposed for including some contributions
from the six-point vertex \cite{Katanin2004, Salmhofer2004, Eberlein2014};
however, until now it was not known how to do this
in a way that captures all parquet diagrams.
In this work,
we present a multiloop fRG (mfRG) scheme,
which sums up all parquet diagrams
to arbitrary order in the interaction.
We apply it to the XES,
a prototypical fermionic problem
with a logarithmically divergent perturbation theory \cite{Giamarchi2004};
in a related publication \cite{Kugler2017}, we develop the mfRG framework for general models.
The XES allows us to focus on two-particle quantities, as these are solely responsible
for the leading logarithmic divergence \cite{Roulet1969, Nozieres1969},
and exhibits greatly simplified diagrammatics.
In fact, it contains the minimal structure required
to study the complicated interplay between different two-particle channels.
We demonstrate how increasing the number of loops
in mfRG improves
the numerical results w.r.t.\ to the known solution of the
parquet equations \cite{Roulet1969, Nozieres1969, Nozieres1969a}.
We establish the equivalence
of the mfRG flow to the parquet approximation 
by showing that both schemes generate the same number of
diagrams order for order in the interaction 
\cite{SuppInfo}\nocite{Gradshteyn2007, Sloane2017}.
	
%
{\sl Model.}---%
The minimal model for the XES is
defined by the Hamiltonian
\begin{equation}
H = \sum_{\epsilon} \epsilon 
c_{\epsilon}^{\dag} c_{\epsilon}^{\phantom\dag}
+ \epsilon_d d^{\dag} d 
+ U c^{\dag} c d^{\dag} d
\EC
\quad
U > 0
\ED
\label{eq:xray_ham}
\end{equation}
Here,
$d$ and $c_{\epsilon}$ respectively annihilate an electron
from a localized, deep core level ($\epsilon_d < 0$)
or a half-filled conduction band with constant density of states $\rho$,
half-bandwidth $\xi_0$, and chemical potential $\mu=0$,
while 
$c = \sum_{\epsilon} c_{\epsilon}$
annihilates a band electron at the core-level site.
In order to describe optical properties of the system,
one examines the particle-hole susceptibility 
$i\Pi(t) = \langle \mathcal{T} d^{\dag}(t) c(t) c^{\dag}(0) d(0) \rangle$.
It exhibits a power-law divergence 
for frequencies close to the absorption threshold,
as found both by the solution of
parquet equations \cite{Roulet1969, Nozieres1969}
and by an exact one-body approach \cite{Nozieres1969a}.
In the Matsubara formalism,
the bare level propagator reads 
$G^d_{\omega} = 1/(i\omega-\epsilon_d)$,
and, focusing on infrared properties,
we approximate the local band propagator as 
$G^c_{\omega} = - i \pi \rho \, \textrm{sgn} (\omega) \Theta(\xi_0 - |\omega|)$.
The particle-hole susceptibility takes the form
(at a temperature $1/\beta \ll |\epsilon_d|$) 
\begin{equation}
\Pi_{\bar{\omega}}
= 
\frac{\rho}{\alpha(u)} \bigg[ 1 - \Big( \frac{i\bar{\omega}+\epsilon_d}{-\xi_0} \Big)^{-\alpha(u)} \bigg]
\EC
\quad
u = \rho U
\EC
\label{eq:phsuscep}
\end{equation}
where $\alpha(u) = 2u + O(u^2)$
and $\epsilon_d$ is considered as a renormalized threshold.
The corresponding retarded correlation function is obtained by
analytic continuation $i\bar{\omega}\to w+i0^+$,
in which case the summands leading to the power-law
are logarithmically divergent as
$u^n \ln^{n+1}(\xi_0/|w+\epsilon_d|)$.
For imaginary frequencies, however,
the perturbative parameter
is finite, with a maximal value of
$u \ln(\xi_0/|\epsilon_d|) \approx 0.9$,
for our choice of parameters.
Our goal will be to reproduce \ER{eq:phsuscep}
using fRG.
%

%
{\sl Parquet formalism.}---%
The particle-hole susceptibility is fully determined by
the one-particle-irreducible (1PI) 
four-point vertex 
via the following relation
(using the shorthand notation $\Gamma^{(4)}_{\omega, \nu, \bar{\omega}}
= \Gamma^{\bar{d} c \bar{c} d}_{\omega, \bar{\omega}+\omega, \bar{\omega}+\nu, \nu}$
\cite{Kugler2017a}):
\begin{align}
\Pi_{\bar{\omega}}
= 
\frac{1}{\beta} \sum_{\omega} G^d_{\omega} G^c_{\bar{\omega} + \omega}
+ \frac{1}{\beta^2} \sum_{\omega, \nu} G^d_{\omega}  G^c_{\bar{\omega} + \omega} 
\Gamma^{(4)}_{\omega, \nu, \bar{\omega}}
G^d_{\nu} G^c_{\bar{\omega} + \nu} 
\ED
\label{eq:phsuscep-gamma4}
\end{align}
In principle, $G^{c}$ and $G^{d}$ are full propagators. However,
for the XES, electronic 
self-energies do not contribute to the leading 
logarithmic divergence \cite{Roulet1969, Nozieres1969},
and we can restrict ourselves to
bare propagators.
\begin{figure}[t]
\includegraphics[width=.48\textwidth]{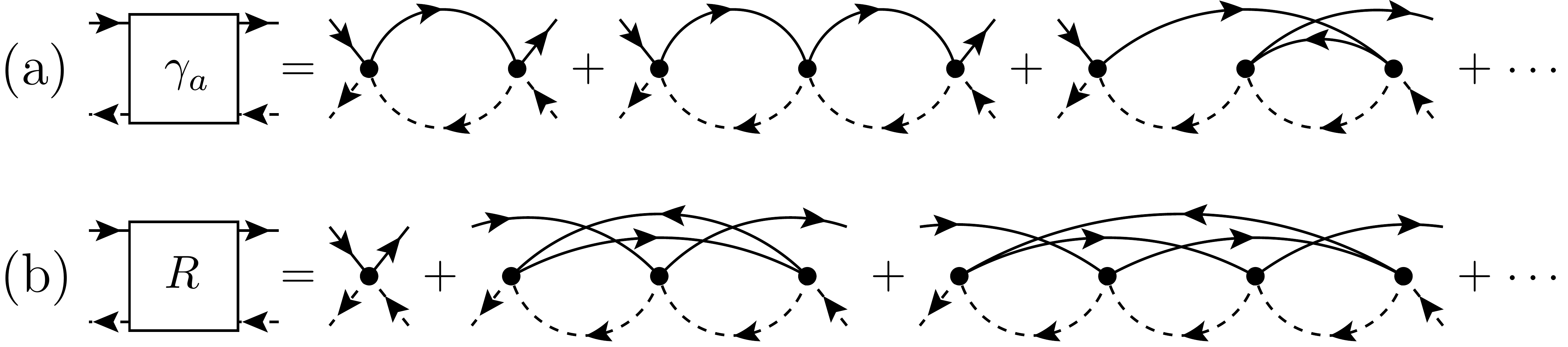}
\caption{%
Low-order diagrams for (a) the vertex reducible in antiparallel lines, $\gamma_{a}$,
and (b) the totally irreducible vertex $R$.
Solid (dashed) lines denote $G^c$ ($G^d$), and a dot the bare vertex $-U$.
The first-order or so-called parquet approximation only retains the bare vertex for $R$.%
}
\label{fig:gamma_ap_and_R}
\end{figure}
Diagrams for the four-point vertex are
exactly classified by the central parquet equation
\begin{equation}
\Gamma^{(4)} 
= 
R + \gamma_{a} + \gamma_{p}
\EC \quad
I_{a} 
= R + \gamma_{p}
\EC 
\quad
I_{p} = R + \gamma_{a}
\ED
\label{eq:parqueteq}
\end{equation}
The leading divergence of the XES 
is determined by only two two-particle channels \cite{Roulet1969, Nozieres1969}:
$\gamma_{a}$ (cf.~\FR{fig:gamma_ap_and_R}(a) \footnote{Since, in a zero-temperature real-frequency treatment \cite{Roulet1969} of X-ray absorption, $G^d$ is purely advanced, we draw diagrams such that all $G^d$ lines are oriented to the left.}) 
and $\gamma_{p}$ contain diagrams reducible
by cutting two antiparallel or parallel lines, respectively, whereas
$I_{a}$ and $I_{p}$ contain diagrams irreducible in the respective channel.
The totally irreducible vertex $R$ [cf.~\FR{fig:gamma_ap_and_R}(b)]
is the only input into
the parquet equations, as the reducible vertices are determined self-consistently
via Bethe-Salpeter equations [cf.~\FR{fig:bs_and_flow}(a)].
Similarly as for the self-energy,
terms of $R$ beyond the bare interaction
only contribute subleadingly
to the XES
and can hence be neglected \cite{Roulet1969, Nozieres1969}.
In this (parquet) approximation, \ER{eq:parqueteq} together with the Bethe-Salpeter
equations for reducible vertices [\FR{fig:bs_and_flow}(a)]
form a closed set and can be solved.
The analytic solution, employing logarithmic accuracy,
provides the leading term of the exponent in \ER{eq:phsuscep}.
Our numerical solution, to which we compare all following results,
is both consistent with the power-law-like behavior 
of \ER{eq:phsuscep} for small frequencies [cf. \FR{fig:regulators_and_parquet}(c)]
and with the corresponding exponent $\alpha(u)$ [cf. \FR{fig:regulators_and_parquet}(d)].
\begin{figure}[t]
\includegraphics[width=.48\textwidth]{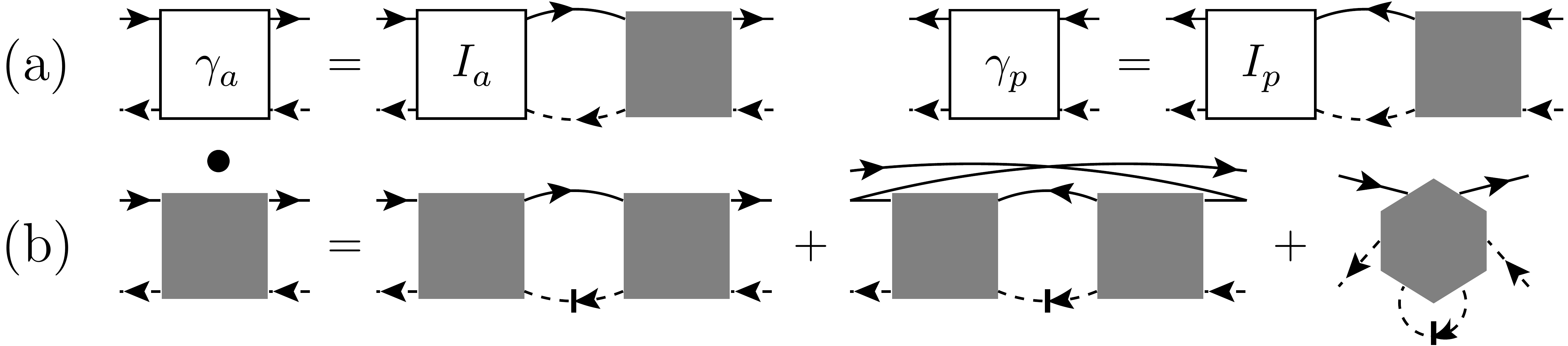}
\caption{%
(a)
Bethe-Salpeter equations in the antiparallel ($a$) and parallel ($p$) channels.
A full square denotes the full vertex $\Gamma^{(4)}$.
(b)
FRG flow equation for both channls relating $\partial_{\Lambda} \Gamma^{(4)}$ 
to $\Gamma^{(4)}$ and $\Gamma^{(6)}$.
The conventional approximation is to set $\Gamma^{(6)}=0$.%
}
\label{fig:bs_and_flow}
\end{figure}
\begin{figure*}[t]
\includegraphics[width=.99\textwidth]{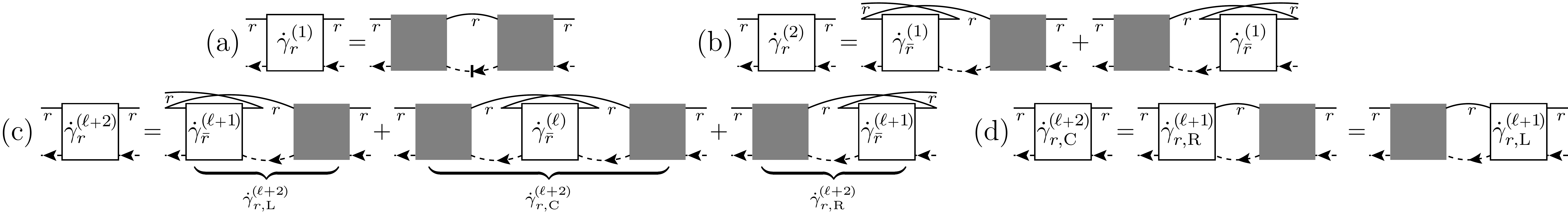}
\vspace{-0.1cm}
\caption{%
Multiloop fRG flow equations,
$\partial_{\Lambda} \gamma_{r} = \sum_{\ell \geq 1} \dot{\gamma}_{r}^{(\ell)}$,
for the four-point vertex reducible in channel $r$, 
with $r=a$ or $p$, and $\bar{r}=p$ or $a$.
The subscript $r$ in the diagrams further symbolizes antiparallel
or parallel $c$-$d$ lines, respectively.
(a)~One-loop, (b)~two-loop, (c)~three- and higher-loop flows.
(d)~One-loop calculation of
$\dot{\gamma}_{r,\textrm{C}}^{(\ell\textrm{\tiny{+}}2)}$,
using the previously computed 
$\dot{\gamma}_{r,\textrm{R}}^{(\ell\textrm{\tiny{+}}1)}$
or
$\dot{\gamma}_{r,\textrm{L}}^{(\ell\textrm{\tiny{+}}1)}$.%
}
\label{fig:multiloop_flow}
\end{figure*}
%

%
{\sl Multiloop fRG flow.}---%
The functional renormalization group provides an
exact flow equation for the four-point vertex
as a function of an RG scale parameter $\Lambda$, serving as infrared cutoff.
Introducing $\Lambda$ only in the bare $d$
propagator, the flow encompassing both channels \cite{SuppInfo}
is illustrated in \FR{fig:bs_and_flow}(b),
where the dashed arrow symbolizes the single-scale propagator 
$S^d_{\Lambda}$.
Neglecting self-energies, we have
$S^d_{\Lambda} = \partial_{\Lambda} G^d_{\Lambda}$,
and $\partial_{\Lambda}\Gamma^{(4)}$ only depends on 
$\Gamma^{(4)}$ and $\Gamma^{(6)}$.
The boundary conditions 
$G^d_{\Lambda_i} = 0$ and
$G^d_{\Lambda_f} = G^d$
imply
$\Gamma^{(4)}_{\Lambda_i} = -U$
and 
$\Gamma^{(6)}_{\Lambda_i} = 0$.
For almost all purposes, it is unfeasible to treat
the six-point vertex exactly. 
Approximations of $\Gamma^{(6)}$ thus render the
fRG flow approximate.
The conventional approximation is to set $\Gamma^{(6)}$ 
and all higher-point vertices to zero,
arguing that they are at least of third order in the interaction.
This affects the resulting four-point vertex starting at third order
and neglects terms that contribute to parquet diagrams \cite{Kugler2017a}. 
Since, however, the parquet approximation involves
\textit{only} four-point vertices, 
it should be possible to encode the influence of six-
and higher-point vertices during the RG flow by four-point
contributions and, still, \textit{fully} capture all parquet graphs.
In the following, we 
show how this can be accomplished using mfRG.
The first observation is that all the diagrammatic content
of the truncated fRG (i.e.~without $\Gamma^{(6)}$)
is two-particle reducible, due to the bubble structure
in the flow equation [first two summands of \FR{fig:bs_and_flow}(b)],
very similar to the Bethe-Salpeter equations [\FR{fig:bs_and_flow}(a)].
The only irreducible contribution is the initial condition of the
vertex, $\Gamma^{(4)}_{\Lambda_i}=-U$. Hence, diagrams generated by the
flow are always of the parquet type.
It is then natural to express $\Gamma^{(4)}$ as follows,
using the channel classification of the parquet equations:
\begin{equation}
\Gamma^{(4)} = -U + \gamma_{a} + \gamma_{p}
\EC \qquad
\partial_{\Lambda} \gamma_{r} = {\textstyle\sum_{\ell\geq1}} \dot{\gamma}_{r}^{(\ell)}
\ED
\end{equation}
Here, $r$ stands for $a$ or $p$ and
$\dot{\gamma}_{r}^{(\ell)}$ 
for diagrams involving
$\ell$ loops connecting full vertices. 
We will show that $\dot{\gamma}_{r}^{(\ell)}$
can be constructed iteratively 
from lower-loop contributions.
The conventional (or \textit{one-loop}) fRG flow in channel $r$ is formulated 
in \FR{fig:multiloop_flow}(a),
where full vertices are connected by an $r$ ``single-scale'' bubble,
i.e., either antiparallel or parallel $G^c$-$S^d$ lines. 
[Detailed diagrams with all arrows and their mathematical 
translations are given in \cite{SuppInfo}, 
Fig.~S2, Eq.~(S2)
.]
If one inserts the bare vertex for $\Gamma^{(4)}$ on the r.h.s.\ of 
such a one-loop flow equation [\FR{fig:multiloop_flow}(a)],
one fully obtains the differentiated second-order vertex.
However, inserting first- and second-order vertices on the r.h.s.\ will
miss some diagrams of the differentiated third-order vertex,
because these invoke an $\bar{r}$ single-scale bubble
that is not generated by $\dot{\gamma}_{r}^{(1)}$
(an overbar denotes the complementary channel:
$\bar{a}=p$, $\bar{p}=a$).
An example of such a missing third-order diagram
is that obtained by differentiating
the rightmost $d$ propagator of the 
third diagram in \FR{fig:gamma_ap_and_R}(a)
(cf.~Fig.~S1 
of \cite{SuppInfo}).
All such neglected contributions can be added to the r.h.s.\ of
the flow equation by hand (replacing bare by full vertices),
resulting in the construction in \FR{fig:multiloop_flow}(b).
It uses an $r$ ``standard'' bubble
[(anti)parallel $G^c$-$G^d$ lines]
to connect the one-loop contribution from
the complementary channel, $\dot{\gamma}_{\bar{r}}^{(1)}$,
with the full vertex, thus generating \textit{two}-loop
contributions.
These corrections have already been suggested from slightly
different approaches \cite{Katanin2004, Eberlein2014}.
The resulting third-order corrected flow will still miss derivatives of 
parquet graphs starting at
fourth order in the interaction.
These can be included via two further additions to the flow,
which have the same form for all higher loop orders,
$\dot{\gamma}_{r}^{(\ell\textrm{\tiny{+}}2)}$
with $\ell \geq 1$ [cf.~\FR{fig:multiloop_flow}(c)].
First, for the flow of $\dot{\gamma}_{r}^{(\ell\textrm{\tiny{+}}2)}$,
an $r$ bubble is used to attach
the previously computed 
$(\ell+1)$-loop contribution from the
complementary channel, $\dot{\gamma}_{\bar{r}}^{(\ell\textrm{\tiny{+}}1)}$,
to either side of the full vertex,
just as in the two-loop case.
Second, by using two $r$ bubbles,
we include the 
differentiated $\ell$-loop vertex from 
the complementary channel, $\dot{\gamma}_{\bar{r}}^{(\ell)}$,
to the flow of $\dot{\gamma}_{r}^{(\ell\textrm{\tiny{+}}2)}$.
Double counting of diagrams in all these contributions
does not occur due to the unique position of the single-scale propagator
\cite{SuppInfo}.
Note that the central term in \FR{fig:multiloop_flow}(c)
can be computed by a one-loop integral, too,
using the previous computations from the same channel, as
shown in \FR{fig:multiloop_flow}(d).
Consequently, the numerical effort in the multiloop corrections
scales linearly in $\ell$.
By its diagrammatic construction,
organized by the number of loops connecting full vertices,
the mfRG flow incorporates \textit{all} differentiated diagrams
of a vertex reducible in channel $r$,
built up from the bare interaction,
and thus captures \textit{all} parquet graphs of the 
full four-point vertex.
Indeed, in \cite{SuppInfo}, we prove algebraically for the XES
that the number of differentiated diagrams in mfRG
matches precisely the number of differentiated parquet graphs.
An $\ell$-loop fRG flow
generates \textit{all} parquet diagrams up to order $n=\ell+1$ in the interaction
and, naturally, generates an increasing number
of parquet contributions at arbitrarily large orders in $U$.
%

%
{\sl Numerical results}---%
In \FR{fig:multiloop_numerics}, we show numerical
results for the XES particle-hole susceptibility.
Using four different regulators (see below),
we compare the susceptibility obtained from an
$\ell$-loop fRG flow to the numerical solution of the
parquet equations.
We find that the one-loop curves differ among each other
and deviate strongly from the parquet result.
With increasing loop order $\ell$,
the multiloop results from all regulators
oscillate around and approach the
parquet result, with very good
agreement already for $\ell=4$.
For $\ell \geq 7$, the oscillations in the relative deviation
(at $\bar{\omega}=0$)
are damped to $\lesssim 2 \%$ (insets, solid line).
A similar behavior is observed for the identity \cite{Wentzell2016}
$\Pi_{\bar{\omega}} = \lim_{|\omega|,|\nu|\to\infty}
\gamma_{a;\omega,\nu,\bar{\omega}}/U^2$
($\bar{\omega}$ is the exchange frequency, and $\omega$, $\nu$ are two 
fermionic frequencies),
which the parquet solution is guaranteed to fulfill
(cf.\ Ref.\ \onlinecite{SuppInfo}, Eq.~(S4) 
and following)
(insets, dashed line).
As regulators, we choose the Litim regulator \cite{Litim2001},
and propagators of the type 
$G^d_{\Lambda}(\omega) = \theta(\omega/\Lambda-1)G^d(\omega)$,
where $\theta(x)$ is either a sharp,
smooth, or oscillating step function 
(cf.~Fig.~4
(a,b); Eq.~(S8) 
of \cite{SuppInfo}).
The fact that different regulators give the same result
in the mfRG flow is a strong indication
for an exact resummation of diagrams.
Let us note that
the mfRG flow also increases the stability
of the solution towards larger interaction.
Whereas, in the one-loop scheme, the four-point vertex
diverges for $u > 0.4$,
higher-loop schemes converge up to larger values of $u$.
The reason is that the one-loop scheme
contains the \textit{full} ladder series of diagrams (in any channel), 
but only \textit{parts} of nonladder diagrams. 
Whereas the (imaginary-frequency) pure particle-hole ladder
already diverges at $u \sim 0.3$, higher-loop extensions
approaching the parquet summation are needed for the full
feedback between both channels to eliminate the divergence.
The equivalence between the
mfRG flow and parquet summation
allows us to explain how the quality of fRG results
depends on the choice of regulator.
Whereas the one-loop scheme only involves
a single-scale bubble
$\Pi_0^S = \sum G^c S^d$,
all extensions invoke successive standard bubbles
$\Pi_0^G = \sum G^c G^d$.
By minimizing the weight of $\Pi_0^G$ compared to $\Pi_0^S$,
one minimizes the effect of the multiloop corrections
and thus the difference between low-level mfRG and parquet.
Indeed, from
\FR{fig:regulators_and_parquet}(a,b) we see
that a regulator with small (large) weight in
$\Pi_0^G$
and large (small) weight in 
$\Pi_0^S$,
such as the oscillating-step (Litim) regulator,
gives comparatively good (bad) agreement with parquet for low $\ell$.
Accordingly, the sharp-step regulator performs slightly
better than its smooth counterpart.
\begin{figure}[t]
\includegraphics[width=.49\textwidth]{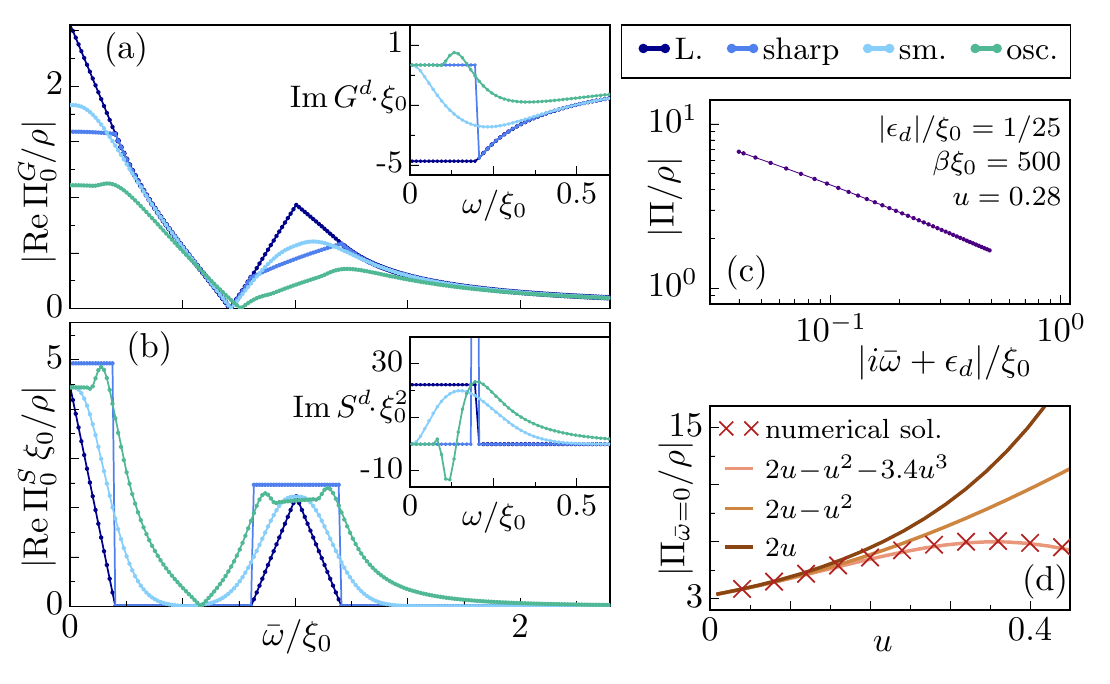}
\caption{%
(a) Noninteracting ``standard'' particle-hole bubble $\Pi_0^G$
and propagator $G^d$ (inset) for 
different regulators [cf.\ \ER{eq:regulator_details} 
of Ref.\ \onlinecite{SuppInfo}] and $\Lambda / \xi_0 = 0.2$.
(b) Same as (a) for the ``single-scale'' bubble $\Pi_0^S$ 
and propagator $S^d$.
(c) Double-logarithmic plot for the particle-hole susceptibility $\Pi$,
obtained from solving the parquet equations.
(d)
$\Pi_{\bar{\omega}=0}(u)$
computed via the parquet equations [$\epsilon_d$, $\beta$ as in (c)]
and according to \ER{eq:phsuscep} with different choices for $\alpha(u)$.
The comparison between these guide-to-the-eye lines and the numerical solution 
confirms that $\alpha(u) \approx 2u$, 
but also shows that subleading contributions become
sizable for larger $u$. These are present since internal numerical
calculations go beyond logarithmic accuracy.%
}
\label{fig:regulators_and_parquet}
\end{figure}
%

%
{\sl Generalizations.}---%
The mfRG flow can be readily extended to more general models,
where one normally does not treat two particle
species separately,
as done here for $c$ and $d$ electrons.
If three two-particle channels 
(antiparallel, parallel, and transverse)
are involved,
the higher-loop flow must incorporate feedback from both
complementary 
channels via
$\dot{\gamma}_{\bar{r}}^{\ell} = \sum_{r' \neq r} \dot{\gamma}_{r'}^{\ell}$
\cite{Kugler2017}.
The self-energy $\Sigma$
enters the $\Gamma^{(4)}$ flow via
full propagators, and,
in the one-loop flow of the four-point vertex [\FR{fig:multiloop_flow}(a)], 
one should follow the usual practice \cite{Katanin2004, Metzner2012}
of using the derivative of the full propagator ($\partial_{\Lambda} G_{\Lambda}$)
instead of the single-scale propagator
($S_{\Lambda} = \partial_{\Lambda} G_{\Lambda}|_{\Sigma=\textrm{const.}}$)
which excludes any differentiated self-energy contributions.
The reason is that,
in the exact fRG flow equation [\FR{fig:bs_and_flow}(b)], 
those diagrams of $\partial_{\Lambda} \Gamma^{(4)}$
that involve 
$\partial_{\Lambda} \Sigma$
are encoded in the six-point vertex.
Evidently, an improved flow for $\Gamma^{(4)}$ 
also improves fRG calculations of the self-energy.
In the parquet formalism, $\Sigma$ is constructed
from the four-point vertex by an exact, self-consistent
Schwinger-Dyson equation \cite{Bickers2004}.
In order to obtain the same self-energy diagrams from
the (in principle) \textit{exact} fRG flow equation for $\Sigma$,
with only the 
vertex in the parquet \textit{approximation}
at one's disposal,
multiloop extensions to the self-energy flow, similar to 
those introduced here, can be performed \cite{Kugler2017}.
Given the self-energy, all arguments about capturing parquet diagrams
(which now consist of dressed lines)
with the multiloop fRG flow remain valid 
since they only involve generic,
model-independent statements about the structure
of two-particle diagrams.
\begin{figure}[t]
\includegraphics[width=.49\textwidth]{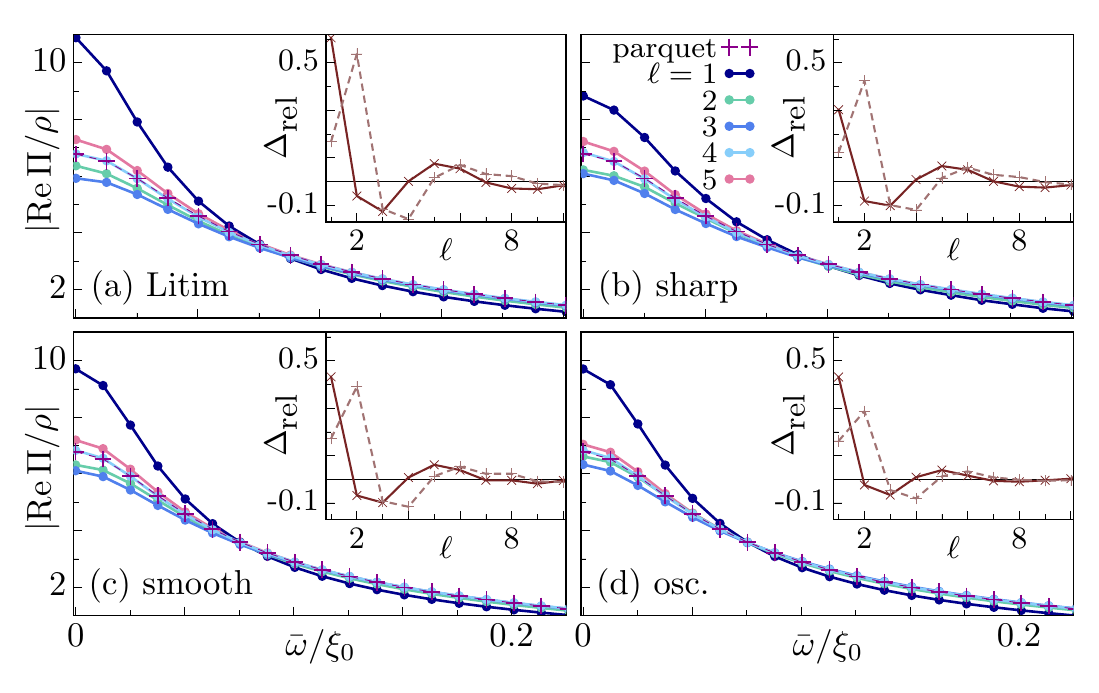}
\caption{%
Numerical solutions for the particle-hole susceptibility $\Pi$,
obtained from the parquet equations and from mfRG with 
different regulators 
[cf.~\FR{fig:regulators_and_parquet}(a,b)],
using the parameters of \FR{fig:regulators_and_parquet}(c).
Insets: relative deviation between 
parquet and mfRG results for $\Pi$ (solid line)
and between $\Pi$ and 
$\lim_{|\omega|,|\nu|\to\infty} \gamma_{a}/U^2$ (dashed line),
all evaluated at $\bar{\omega}=0$.%
}
\label{fig:multiloop_numerics}
\end{figure}
The mfRG flow is applicable for any initial condition $\Gamma^{(4)}_{\Lambda_i}$.
An example where one would not start from $G_{\Lambda_i}=0$,
as done here, arises in the context of
dynamical mean-field theory (DMFT) \cite{Georges1996}.
There, the goal of adding nonlocal correlations, 
with the local vertex from DMFT ($\Gamma^{(4)}_{\textrm{DMFT}}$)
as input, can be pursued using fRG \cite{Taranto2014}.
Alternatively, this goal is 
also being addressed by
using the parquet equations in the 
dynamical vertex approximation 
(D$\Gamma$A) 
\cite{Toschi2007, Held2008, Valli2010}.
However, the latter approach 
requires the \textit{diagrammatic}
decomposition of the \textit{nonperturbative} vertex
\footnote{Alternatives to D$\Gamma$A which do not require the totally irreducible vertex are the dual fermion \cite{Rubtsov2008,*Brener2008,*Hafermann2009} and the related 1PI approach \cite{Rohringer2013}. However, upon transformation to the dual variables, the bare action contains $n$-particle vertices for all $n$. Recent studies \cite{Ribic2017,*Ribic2017a}
show that the corresponding six-point vertex yields sizable contributions for the (physical) self-energy, and it remains unclear how a truncation in the (dual) bare action can be justified.}
$\Gamma^{(4)}_{\textrm{DMFT}} = R + \sum_r \gamma_r$,
which yields diverging results close
to a quantum phase transition 
\nocite{Ribic2017} \nocite{Ribic2017a}
\cite{Schaefer2013,Schaefer2016,*Gunnarsson2017}.
In contrast, the mfRG flow is built from the \textit{full} vertex 
$\Gamma^{(4)}_{\textrm{DMFT}}$ and could thus 
be used to scan a larger region of the phase diagram.
%

%
{\sl Conclusion.}---%
Using the X-ray-edge singularity as an example,
we have presented multiloop fRG flow equations,
which sum up all parquet diagrams to arbitrary order,
so that solving the mfRG flow is equivalent
to solving the (first-order) parquet equations.
Our numerical results demonstrate that
solutions of an $\ell$-loop
flow quickly approach the parquet result
with increasing $\ell$.
This applies for a variety of regulators,
confirming an exact resummation of diagrams.
The mfRG construction is generic and can 
be readily generalized to more complex models.
The mfRG-parquet equivalence established here
shows that one-loop fRG calculations generate 
only a subset of (differentiated) parquet diagrams
and that a multiloop fRG flow is needed to 
reproduce parquet results.
From a practical point of view,
mfRG appears advantageous over solving the
parquet equations since solving a first-order
ordinary differential equation is numerically
more stable than solving a self-consistent equation.
Moreover, one can choose a suitable regulator and
flow from any initial action.
Altogether, the mfRG scheme 
achieves, in effect, a solution of the 
(first-order) parquet equations
while retaining all treasured fRG advantages: 
no need to solve self-consistent equations, purely one-loop costs, and
freedom of choice for regulators.
We thank A.\ Eberlein, C.\ Honerkamp, S.\ Jakobs, V.\ Meden, W.\ Metzner,
and A.\ Toschi for useful discussions
and acknowledge support by the Cluster of Excellence
Nanosystems Initiative Munich.
F.B.K.\ acknowledges funding from
the research school IMPRS-QST.
\bibliographystyle{apsrev4-1}
\bibliography{references}

\newpage

\begin{center}
{\bfseries\large Supplemental material}\\~\\
\end{center}
This supplement consists of four parts.
First, we show detailed equations
for the mfRG flow, the identity between susceptibility
and reducible vertex, and the regulators we used.
Second, we provide the numerical details of our computations.
Third, we prove algebraically for the XES that the mfRG flow 
generates all parquet diagrams at arbitrary order, 
based on expanding the parquet and flow equations in the interaction
and counting diagrams.
Last, we briefly mention that
many quantities appearing in this proof
happen to have an interpretation as
giving the number of special
paths on a triangular grid.
\section*{S-I. Detailed equations}
Figure \FRn{fig:two-loop_example} illustrates how the
two-loop corrections of mfRG cure the flow of the vertex
$\gamma_{a}$ at third order in the interaction.
Figure \FRn{fig:multiloop_gamma} shows the detailed form of 
the mfRG flow equations, corresponding to 
\FR{fig:multiloop_flow}.
Figure \FRn{fig:two-loop_example} illustrates how the
two-loop corrections of mfRG cure the flow of the vertex
$\gamma_{a}$ at third order in the interaction.
Figure \FRn{fig:multiloop_gamma} shows the detailed form of 
the mfRG flow equations from \FR{fig:multiloop_flow}.
In principle \cite{SuppInfo},
the flow equations also contain contributions from a third (transversal)
channel, where the \textit{interband} vertex $\Gamma^{\bar{d}c\bar{c}d}$ is connected to an \textit{intraband} vertex $\Gamma^{\bar{d}d\bar{d}d}$ by valence band lines $G^d$ and $S^d$.
However, one can easily see that, for the XES, all such terms contribute subleadingly and belong to higher-order diagrams of $R$ in the parquet treatment \cite{Roulet1969}. 
Hence, they are neglected throughout this work.
The mathematical translation of our flow equations
only requires the formula for an $r$
bubble
connecting two vertices
(where $r=a,p$).
This is most compactly written in a notation
adapted to the respective channel:
The three independent frequencies necessary to
describe a full vertex can be chosen to include
two fermionic frequencies combined with either
the bosonic \textit{exchange} frequency $\bar{\omega}_a$,
suited for the antiparallel channel,
or the bosonic \textit{pairing} frequency $\bar{\omega}_p$,
suited for the parallel channel.
This is, however, merely a choice of parametrization
and does not require any properties of the vertex itself.
We choose the parametrization according to
\begin{minipage}{.3\textwidth}
\centering
\begin{supp-subalign}[eq:channel_notation]
\nonumber \\
\mathcal{V}_{\omega, \nu, \bar{\omega}_a} & = 
\mathcal{V}^{\bar{d}c\bar{c}d}_{\omega, \bar{\omega}_a+\omega, \bar{\omega}_a+\nu, \nu}
\EC \\
\nonumber \\
\mathcal{V}_{\omega, \nu, \bar{\omega}_p} & = 
\mathcal{V}^{\bar{d}c\bar{c}d}_{\omega, \bar{\omega}_p-\nu, \bar{\omega}_p-\omega, \nu}
\EC
\end{supp-subalign}%
\end{minipage}%
\begin{minipage}{.18\textwidth}
\centering
\includegraphics[width=0.6\textwidth]{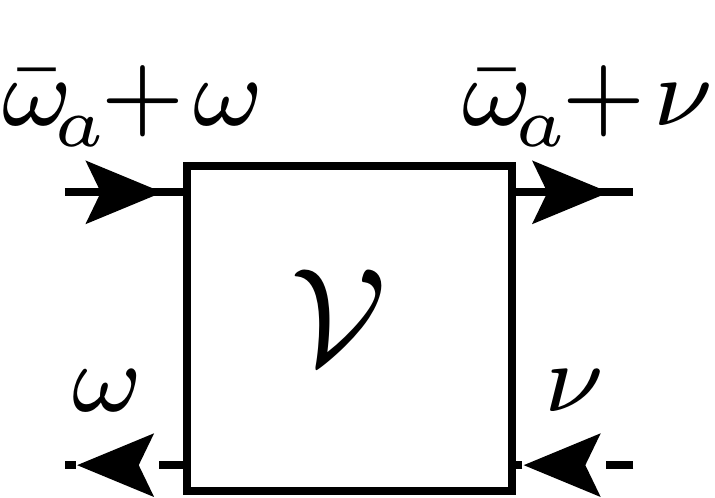} \\
\includegraphics[width=0.6\textwidth]{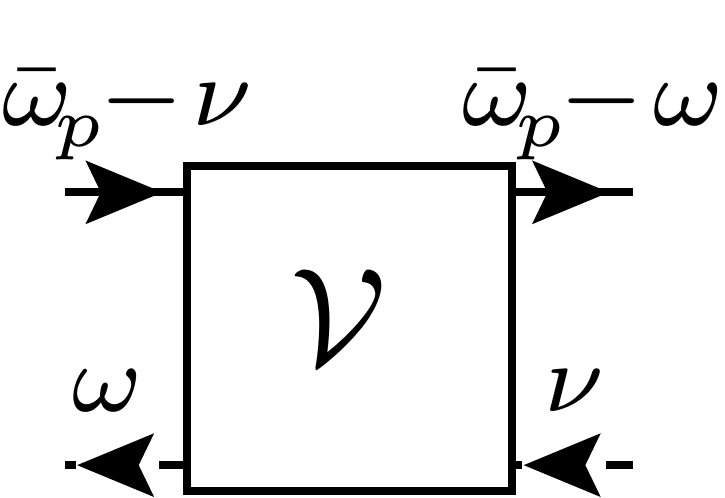}
\end{minipage}
where the bosonic frequencies are related via
$\bar{\omega}_p = \bar{\omega}_a + \omega + \nu$.

In this notation, an $r$ bubble $\mathcal{V}_{r}$
connecting the vertices
$\mathcal{V}^{'}$ and $\mathcal{V}^{''}$
can be computed as follows:
\begin{supp-equation}
\mathcal{V}_{r; \omega, \nu, \bar{\omega}_r}
= \frac{1}{\beta} \sum_{\omega'}
\mathcal{V}^{'}_{\omega, \omega', \bar{\omega}_r}
G^d_{\omega'} G^c_{\bar{\omega}_r + \sigma_r \omega'} 
\mathcal{V}^{''}_{\omega', \nu, \bar{\omega}_r}
\EC
\label{eq:rbubble}
\end{supp-equation}%
with $\sigma_a = 1$ and $\sigma_p = -1$.
The channel notation \ERn{eq:channel_notation} is 
also used in the identity
between particle-hole susceptibility $\Pi$ and reducible vertex
$\gamma_a$ considered in \FR{fig:multiloop_numerics}.
If we, more generally, denote the susceptibility in the antiparallel channel by
$\Pi_a = \Pi$ and the one in the parallel channel by $\Pi_p$, 
the relation between susceptibility and 1PI vertex,
already used in \ER{eq:phsuscep-gamma4}, reads
\begin{supp-equation}
\Pi_{r;\bar{\omega}_r}
= 
\frac{1}{\beta} \sum_{\omega} G^d_{\omega} G^c_{\bar{\omega}_r + \sigma_r \omega} 
\Big( 1 + 
\frac{1}{\beta} \sum_{\nu} 
\Gamma^{(4)}_{\omega, \nu, \bar{\omega}_r}
G^d_{\nu} G^c_{\bar{\omega}_r + \sigma_r \nu} 
\Big)
\ED
\label{eq:phsuscep-gamma4_general}
\end{supp-equation}%
The identity between susceptibility and reducible vertex
\cite{Wentzell2016} is given by
\begin{supp-equation}
\lim_{|\omega|, |\nu| \to \infty} \gamma_{r; \omega, \nu, \bar{\omega}_r} =
U^2 \Pi_{r; \bar{\omega}_r}
\ED
\label{eq:pi_gamma_id}
\end{supp-equation}%
\begin{supp-figure}[t]
\includegraphics[width=0.48\textwidth]{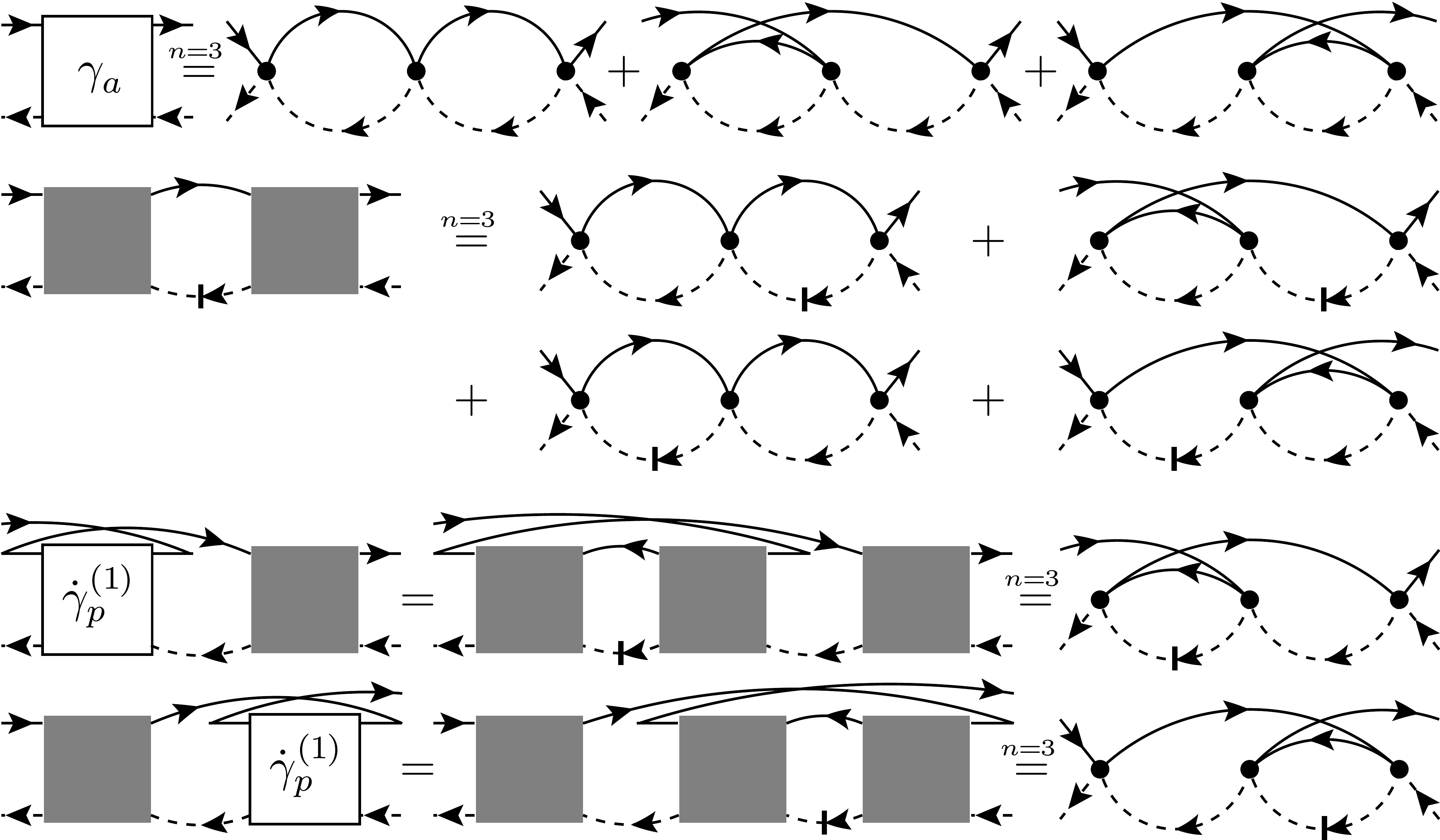}
\caption{%
First row: All third-order contributions to $\gamma_a$.
Its flow is described by the six diagrams obtained by
differentiating each dashed line once. In the mfRG scheme,
these six diagrams are encoded in 
$\dot{\gamma}_a^{(1)}$ (second and third rows)
and $\dot{\gamma}_a^{(2)}$ (last two rows),
the one- and two-loop flow equations
[cf.~\FR{fig:multiloop_gamma}] for
$\gamma_a$, respectively.
The third-order contributions are obtained by
inserting first- and second-order diagrams for the full vertex.%
}
\label{fig:two-loop_example}
\end{supp-figure}
To see that a solution of the parquet equations
with any approximation for the totally irreducible vertex $R$
is guaranteed to fulfill \ER{eq:pi_gamma_id}, we note
first that, by the very fact that $R$ is totally irreducible,
we have
\begin{supp-equation}
\lim_{|\omega| \to \infty} R_{\omega, \nu, \bar{\omega}_r} = -U
\ED
\label{eq:limR}
\end{supp-equation}%
Regarding the reducible vertices, we can perform the limit
in the Bethe-Salpeter equations [\FR{fig:bs_and_flow}(a)]
and obtain
\begin{supp-subalign}[eq:limAP]
\lim_{|\omega| \to \infty} \gamma_{\bar{r}; \omega, \nu, \bar{\omega}_r} 
& = 0
\EC \quad \Rightarrow
\lim_{|\omega| \to \infty} I_{r; \omega, \nu, \bar{\omega}_r} = -U
\EC 
\label{eq:limP} \\
\lim_{|\omega| \to \infty} \gamma_{r; \omega, \nu, \bar{\omega}_r}
& =
-\frac{U}{\beta} \sum_{\omega'}
G^d_{\omega'} G^c_{\bar{\omega}_r + \sigma_r \omega'} 
\Gamma^{(4)}_{\omega', \nu, \bar{\omega}_r}
\ED
\label{eq:limA}
\end{supp-subalign}%

\begin{supp-starfigure}[t]
\includegraphics[width=0.99\textwidth]{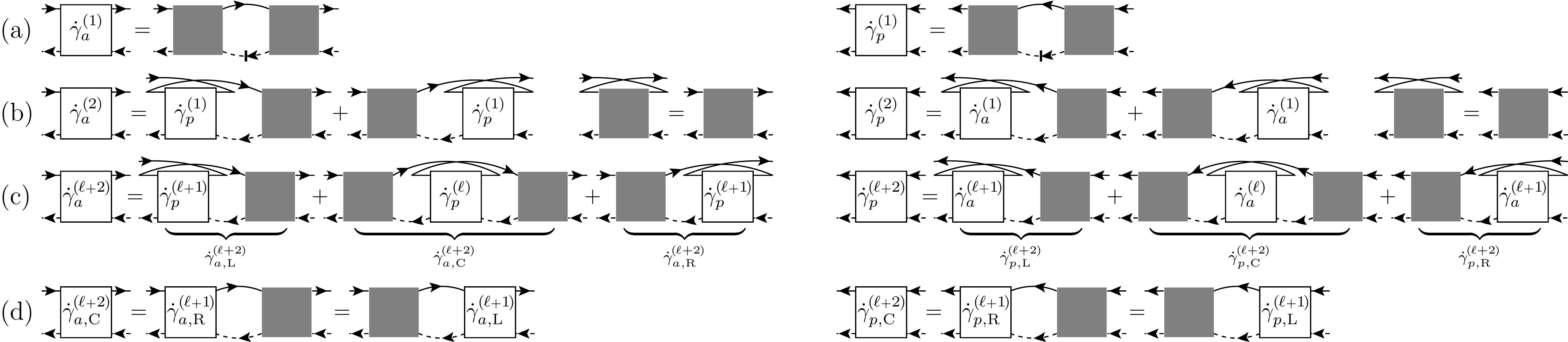}
\caption{%
Multiloop flow equations in the (left) antiparallel and
(right) parallel channels, corresponding to \FR{fig:multiloop_flow}.%
}
\label{fig:multiloop_gamma}
\end{supp-starfigure}
By symmetry [cf.~\ER{eq:vertex_symmetries}],
\EsR{eq:limR}, \ERn{eq:limAP} also hold for $\omega \leftrightarrow \nu$,
and we further deduce
\begin{supp-align}
\lim_{|\nu| \to \infty} \Gamma^{(4)}_{\omega', \nu, \bar{\omega}_r} 
& =
\lim_{|\nu| \to \infty} R_{\omega', \nu, \bar{\omega}_r} +
\lim_{|\nu| \to \infty} \gamma_{r; \omega', \nu, \bar{\omega}_r} \nonumber \\
& = -U
-\frac{U}{\beta} \sum_{\omega''}
\Gamma^{(4)}_{\omega', \omega'', \bar{\omega}_r}
G^d_{\omega''} G^c_{\bar{\omega}_r + \sigma_r\omega''}
\ED
\label{eq:limG}
\end{supp-align}%
Adding the limit $\lim_{|\nu|\to\infty}$ to \ER{eq:limA} and using \EsR{eq:phsuscep-gamma4_general}
and \ERn{eq:limG}
yields the identity \ERn{eq:pi_gamma_id}.
Next, we give the mathematical definition of the regulators,
which we have used in the numerical calculations [\FR{fig:multiloop_numerics}]
and already illustrated in \FR{fig:regulators_and_parquet}(a,b):
\begin{supp-subalign}[eq:regulator_details]
G^{d, \textrm{L}}_{\omega, \Lambda} 
& = 
\frac{1}{i \,\textrm{sgn}(\omega) \,\textrm{max}(|\omega|, \Lambda) - \epsilon_d} 
\label{eq:reg-Litim}
\EC \\
G^{d, \textrm{sharp}}_{\omega, \Lambda} 
& = 
\Theta(|\omega| - \Lambda) \frac{1}{i \omega - \epsilon_d} 
\label{eq:reg-sharp}
\EC \\
G^{d, \textrm{smooth}}_{\omega, \Lambda} 
& = 
\bigg[ 1 - \,\mathrm{e}^{ - \big( \frac{|\omega|}{\Lambda} \big)^{a} } \bigg]
\frac{1}{i \omega - \epsilon_d} 
\EC \quad a = 2
\label{eq:reg-smooth}
\EC \\
G^{d, \textrm{osc.}}_{\omega, \Lambda} 
& = 
\,\mathrm{e}^{ - \big( \frac{\Lambda}{|\omega|} \big)^{a} \big[1 - i b \,\textrm{sgn}(w) \big] }
\frac{1}{i \omega - \epsilon_d} 
\EC \quad a=2\EC \ b=1
\ED
\label{eq:reg-osc}
\end{supp-subalign}%
The regulator in \ER{eq:reg-Litim} is known as Litim regulator \cite{Litim2001}.
Note that the parameters in \EsR{eq:reg-smooth} and \ERn{eq:reg-osc}, $a>0$ and $b$, can also be chosen differently,
keeping the boundary conditions $G^d_{\Lambda_i=\infty} = 0$ and
$G^d_{\Lambda_f=0} = G^d$ fulfilled.
Finally, we remark that, in principle,
the band gap is the largest energy scale in the XES.
This would require $|\epsilon_d| \gg \xi_0$.
However, in the choice of the Hamiltonian [\ER{eq:xray_ham}],
we have already restricted ourselves to an interband density-density
interaction, which implies
individual particle-number conservation.
As a consequence, we are free to choose \textit{any} 
numerical value for $\epsilon_d$,
the only exception being $\epsilon_d=0$, which violates analytic properties
of the (bare) susceptibility \cite{Kugler2017a}.
In fact, we find small values for $|\epsilon_d|$ most suitable to
visualize the power-law divergence in the
particle-hole susceptibility for imaginary frequencies [cf.~\ER{eq:phsuscep}].
\section*{S-II. Numerical details}
We have solved the self-consistent parquet equations 
[\ER{eq:parqueteq}, \FR{fig:bs_and_flow}(a)] by an iterative algorithm.
For that, we use the initial values $\gamma_r = 0$ and
an update rule that combines
the previous value and the
predicted value from the Bethe-Salpeter equations according to 
\begin{supp-equation}
\gamma_r^{\textrm{new}} = z \gamma_r^{\textrm{pred.}}
+ (1-z) \gamma_r^{\textrm{prev.}}
\EC \quad
z \lesssim 0.2
\ED
\end{supp-equation}%
The mfRG flow equations are solved by an adaptive-step Runge-Kutta algorithm.
The numerical costs of the mfRG flow and the parquet algorithm are similar: In both scenarios, one computes bubbles of vertices multiple times---either to evaluate the flow equations during the mfRG flow or to evaluate the Bethe-Salpeter equations during a self-consistency loop in the parquet algorithm.
In either case,
we use a parametrization of four-point vertices which 
accounts for the important high-frequency asymptotics \cite{Li2016,Wentzell2016}.
This parametrization \cite{Wentzell2016} is adapted to the channel
in which a vertex is reducible:
We approximate the frequency dependence of a vertex
reducible in channel $r$,
using the respective channel notation 
from \ER{eq:channel_notation}, 
by
\begin{supp-align}
\label{eq:vertex_parametrization}
\gamma_{r; \omega, \nu, \bar{\omega}_r} & =
\Theta(\Omega_1 - |\bar{\omega}_r|) K^1_{\bar{\omega}_r} \\ 
& \ +
\Theta(\Omega_2 - |\bar{\omega}_r|) \Theta(\Omega_2 - |\omega|) 
K^2_{\bar{\omega}_r, \omega} \nonumber \\ 
& \ +
\Theta(\Omega_2 - |\bar{\omega}_r|) \Theta(\Omega_2 - |\nu|) 
\bar{K}^2_{\bar{\omega}_r, \nu} \nonumber \\ 
& \ +
\Theta(\Omega_3 - |\bar{\omega}_r|) \Theta(\Omega_3 - |\omega|)  \Theta(\Omega_3 - |\nu|) 
K^3_{\bar{\omega}_r, \omega, \nu}
\ED \nonumber 
\end{supp-align}%
Note that the first summand in this parametrization 
already incorporates the limit used in \ER{eq:pi_gamma_id}.
We have chosen the cutoffs $\Omega_i$ in \ER{eq:vertex_parametrization}
such that we keep
$1000$, $500$, and $100$ positive frequencies
on each axis
for $K^1$, $K^2$ and $\bar{K}^2$, and $K^3$, respectively.
Using the symmetries for vertices \cite{Wentzell2016},
\begin{supp-equation}
(\mathcal{V}_{\omega, \nu, \bar{\omega}_r})^{*} =
\mathcal{V}_{-\omega, -\nu, -\bar{\omega}_r}
\EC \quad
\mathcal{V}_{\omega, \nu, \bar{\omega}_r} = 
\mathcal{V}_{\nu, \omega, \bar{\omega}_r}
\EC
\label{eq:vertex_symmetries}
\end{supp-equation}%
further reduces the computational effort.
Note that, while the latter symmetry holds for
$\gamma^{(\ell)}_{r}$ and $\gamma^{(\ell)}_{r,\textrm{C}}$,
it does not hold for
$\gamma^{(\ell)}_{r,\textrm{L}}$ and $\gamma^{(\ell)}_{r,\textrm{R}}$ individually.
Instead, one has 
$\gamma^{(\ell)}_{r,\textrm{L}; \omega,\nu,\bar{\omega}_r}
= \gamma^{(\ell)}_{r,\textrm{R}; \nu,\omega,\bar{\omega}_r}$.
The Matsubara summations in all our calculations are naturally
restricted to a finite frequency interval, since we approximate the
$c$ propagator using a sharp cutoff:
\begin{supp-align}
G^c_{\omega} & = \rho \int_{-\xi_0}^{\xi_0} \textrm{d}\epsilon
\frac{1}{i\omega - \epsilon} = 
-2i\rho \arctan \Big( \frac{\xi_0}{\omega} \Big) \nonumber \\
& = -i \pi \rho \,\textrm{sgn}(w) \Theta(\xi_0 - |\omega|)
+ O \Big( \frac{\xi_0}{\omega} \Big)
\ED
\end{supp-align}%
At an inverse temperature of $\beta \xi_0 = 500$, this yields
about $160$ summands.
\section*{S-III. Proof of equivalence}
We prove below for the XES that solving the full mfRG flow
is equivalent to solving the (first-order) parquet equations. 
We also show that a solution of an $\ell$-loop fRG flow
fully contains all parquet graphs up to order $n=\ell+1$.
In order to check that the parquet vertex is a solution
of the mfRG flow equation (viz., an ordinary differential equation),
one has to verify that the initial condition is fulfilled 
and that the differential equation is fulfilled (during the whole flow). 
At the initial scale 
($\Lambda_i=\infty$, $G^d_{\Lambda_i}=0$, $\Gamma^{(4)}_{\Lambda_i}=-U$)
the parquet vertex is trivially given by the bare vertex; thus the initial condition is fulfilled. 
At an arbitrary scale parameter $\Lambda$ during the flow,
inserting all parquet diagrams for the vertex into, e.g.,
the one-loop flow equation 
generates only a subset of all differentiated parquet diagrams 
(cf.\ \FR{fig:two-loop_example}),
i.e., the differential equation is not fulfilled. 
However, inserting all parquet diagrams into the full mfRG flow equation
yields all differentiated parquet diagrams, 
i.e., the differential equation is fulfilled.
To show that, indeed, all differentiated parquet diagrams are generated in mfRG,
we proceeds in two steps:
First, we argue that, by the structure of the mfRG flow,
the differentiated diagrams are of the parquet type
without any double counting.
Second, we show (without caring about the specific form of a diagram)
that the number of differentiated diagrams in mfRG
exactly matches the number of differentiated parquet graphs 
order for order in the interaction.
\subsection{No double counting in mfRG}
The only totally irreducible contribution to the four-point vertex
contained in the multiloop (or conventionally truncated) fRG flow is the 
bare interaction stemming from the initial condition of the vertex.
All further diagrams on the r.h.s.\ of the flow equations
are obtained by iteratively
combining two vertices with parallel or
antiparallel propagators.
Hence, they correspond to differentiated \textit{parquet} diagrams
in the respective channel.
The fact that there is no double counting in mfRG 
is easily seen employing arguments of diagrammatic reducibility 
and the unique position of the single-scale propagator in differentiated diagrams.
To be specific, let us consider here the channel reducible 
in antiparallel lines [cf.~left side of \FR{fig:multiloop_gamma}]; 
the arguments for the other channel are completely analogous.
First, we note that diagrams in the one-loop term always differ from higher-loop ones.
The reason is that, in higher-loop terms, 
the single-scale propagator appears in the vertex coming from 
$\partial_{\Lambda} \gamma_{p}$. 
This can never contain vertices connected by an antiparallel
$G^c$-$S^d$ bubble, since such terms 
only originate upon differentiating $\gamma_{a}$.
Second, diagrams in the left, center, or right part of an 
$\ell$-loop contribution always differ.
This is because the vertex $\gamma^{(\ell)}_{p}$
is irreducible in antiparallel lines.
The left part is then reducible in antiparallel lines \textit{only after} the single-scale propagator appeared,
the right part \textit{only before},
and the center part is reducible in this channel
\textit{before and after} $S^d$.
Third, the same parts (say, the left parts) of different 
loop contributions ($\ell \neq \ell'$) are always different.
Assume they agreed:
As the antiparallel bubble induces the first (leftmost) reducibility 
in this channel, already
$\gamma^{(\ell)}_{p}$ and $\gamma^{(\ell')}_{p}$ 
would have to agree.
For these, only the same parts can agree, as mentioned before.
The argument then proceeds iteratively until one compares 
the one-loop part to a higher-loop ($|\ell - \ell'| + 1$) one.
These are, however, distinct according to the first point.

To summarize: All mfRG diagrams belong to the parquet class
and are included at most once. 
To show that \textit{all} differentiated parquet diagrams are included,
it remains to compare their number
to the number of diagrams in mfRG.

\subsection{Counting the number of diagrams}
To count the number of diagrams
generated by the parquet equations and mfRG, 
we expand the parquet (Bethe-Salpeter) and
flow equations in the interaction.
As we need not consider the specific form of a diagram, 
the calculation is identical for both channels.
Let us denote the number of parquet diagrams of $\Gamma^{(4)}$ at
order $n$ by $P_0(n)$ (mnemonic: $P$ for parquet).
A $\Gamma^{(4)}$ diagram of order $n$ contains $n-1$ scale-dependent $d$ lines.
Differentiating an $n$-th order diagram by $\Lambda$ thus produces
$n-1$ differentiated diagrams,
and, in total, we have $P_0(n)(n-1)$ differentiated diagrams.
Let us further denote the number of differentiated diagrams at order $n$
in one channel, generated
by mfRG at loop order $\ell$, by $F_{\ell}(n)$ (mnemonic: $F$ for flow).
The $\ell$-loop contributions start at order $n=\ell+1$
in the interaction,
i.e., $F_{\ell}(n)=0$ for $n \leq \ell$.
To show that all parquet diagrams are generated by the (full) mfRG flow, we thus have to establish the following equality:
\begin{supp-equation}
P_0(n)(n-1) = 2 \sum_{\ell=1}^{\infty} F_{\ell}(n) = 2 \sum_{\ell=1}^{n-1} F_{\ell}(n)
\ED
\label{eq:equiv_toshow}
\end{supp-equation}%
In order to sum the parquet graphs up to order $n$, it
suffices to solve the multiloop fRG flow up to loop order $\ell=n-1$.
First, let us count the number of parquet diagrams.
From the Bethe-Salpeter equations [cf.~\FR{fig:bs_and_flow}(a)], one can directly deduce the
number of diagrams at order $n$ inherent in
$\gamma$ (of any channel), $P_{\gamma}(n)$, 
given the number of diagrams in $I$, $P_{I}$,
and in $\Gamma^{(4)}$, $P_0$:
\begin{supp-equation}
P_{\gamma}(n) = \sum_{m=1}^{n-1} P_{I}(m)P_0(n-m)
\ED
\label{eq:bscount}
\end{supp-equation}%

As both $I$ and $\Gamma$ start at order $1$,
the order on the l.h.s.\ exceeds the maximal order of a diagram on the r.h.s.\
From the parquet equations, we further know
\begin{supp-equation}
P_0(1)=1=P_{I}(1)\ESC \quad
P_0(n)=2P_{\gamma}(n)=2P_{I}(n)
\EC \,
n \geq 2
\ED
\end{supp-equation}%
Inserting this, we obtain a closed relation for $P_0$:
\begin{supp-equation}
P_0(n) = \sum_{m=1}^{n-1} P_0(m)P_0(n-m) + P_0(n-1)
\EC
\quad n \geq 2
\ED
\label{eq:parquet_recursion}
\end{supp-equation}%
Let us solve this recursion by the method of generating functions.
We define the generating function $p_0(x)$ for the sequence $P_0(n)$ by
\begin{supp-equation}
p_0(x) = \sum_{n=1}^{\infty} P_0(n) x^{n-1}
\end{supp-equation}%
and calculate
\begin{supp-align}
x p_0(x)^2 & = x \sum_{n,m=1}^{\infty} P_0(n) P_0(m) x^{n+m-2} \nonumber \\
& = \sum_{n=2}^{\infty} x^{n-1} \sum_{m=1}^{n-1} P_0(m)  P_0(n-m) \nonumber \\ 
& = \sum_{n=2}^{\infty} P_0(n) x^{n-1} - \sum_{n=2}^{\infty} P_0(n-1) x^{n-1} \nonumber \\ 
& = \sum_{n=1}^{\infty} P_0(n) x^{n-1} - 1 - x \sum_{n=1}^{\infty} P_0(n) x^{n-1}
\ED
\end{supp-align}%
From this, we find the defining equation for the generating function,
\begin{supp-equation}
x p_0(x)^2 + (x-1) p_0(x) +1 = 0
\EC
\label{eq:defeq}
\end{supp-equation}%
to which the solution with positive Taylor coefficients is
\begin{supp-equation}
p_0(x) = \frac{1-x-\sqrt{1-6x+x^2}}{2x}
\ED
\end{supp-equation}%
Recognizing that $(1-2tx+x^2)^{-\lambda}$ 
is the generating function 
for Gegenbauer polynomials $C_{n-1}^{\lambda}(t)$
\cite{Gradshteyn2007}, we find
\begin{supp-equation}
P_0(n) = - \frac{1}{2} C_n^{-1/2}(3)
\EC \quad
n \geq 2
\end{supp-equation}%
and can read off $P_0(n)$ from a tabulated sequence:
\begin{supp-equation}
P_0: 1,\, 2,\, 6,\, 22,\, 90,\, 394,\, 1806,\, 8558, \dots
\label{eq:parquet_sequence}
\end{supp-equation}%
Note that $P_0(n)$ grows exponentially for large $n$. 
This is much less than the number of all, i.e., parquet and 
nonparquet diagrams of $\Gamma^{(4)}$, which grows
faster than $n!$.
The defining equation for the generating function \ERn{eq:defeq}
can be used to find the generating function $q(x)$ of the
related sequence $P_0(n)(n-1)$:
\begin{supp-equation}
q(x) = \sum_{n=1}^{\infty} P_0(n) (n-1) x^{n-1} = x 
p'_0(x)
\ED
\end{supp-equation}%
Differentiating \ER{eq:defeq}, we find the expression
\begin{supp-align}
0 & = p_0(x)^2 + p_0(x) + [ 1 - x + 2 x p_0(x) ] 
p'_0(x)
\EC \nonumber \\ 
\Rightarrow \ q(x) & = xp_0(x) \frac{p_0(x)+1}{1-x-2xp_0(x)}
\ED
\end{supp-align}%
Next, we count the number of differentiated diagrams generated by mfRG.
For this purpose, we consider the auxiliary vertices
in \FR{fig:multiloop_proof1},
which can be seen as the building blocks of the
multiloop flow equations (\FR{fig:multiloop_gamma}).
Denoting the number of diagrams of $\dot{\mathcal{V}}_{\ell}$ at order $n$ 
by $P_{\ell}(n)$,
we find, 
given all parquet diagrams in the full vertex $\Gamma^{(4)}$,
similar to \ER{eq:bscount} the relation
\begin{supp-equation}
P_{\ell+1}(n) = \sum_{m=1}^{n-1} P_{\ell}(m)P_{0}(n-m) 
\ED
\label{eq:multiloop_sequence}
\end{supp-equation}%

This convolution of two sequences
can be expressed in terms of the product
of their generating functions, defined by
$p_{\ell}(x) =  \sum_{n=1}^{\infty} P_{\ell}(n) x^{n-1} $:
\begin{supp-align}
x p_{\ell}(x) p_{0}(x) & = x \sum_{n,m=1}^{\infty} P_{\ell}(n) P_{0}(m) x^{n+m-2} \nonumber \\
& = \sum_{n=2}^{\infty} x^{n-1} \sum_{m=1}^{n-1} P_{\ell}(m)  P_{0}(n-m) \nonumber \\ 
& = \sum_{n=2}^{\infty} P_{\ell+1}(n) x^{n-1} = p_{\ell+1}(x)
\ED
\label{eq:multiloop_gf}
\end{supp-align}%
As a direct consequence, we have
\begin{supp-equation}
p_{\ell}(x) = x^{\ell} p_0^{\ell+1}(x)
\ESC \qquad
P_{\ell}(n)=0, \quad \ell \geq n
\ED
\label{eq:aux_genfunc}
\end{supp-equation}%
\begin{supp-figure}[t]
\includegraphics[width=0.48\textwidth]{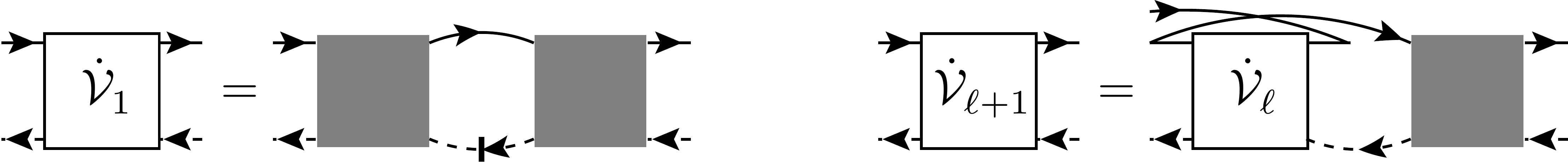}
\caption{%
One-loop equations for auxiliary vertices,
which can be seen as building blocks for the multiloop flow equations
(\FR{fig:multiloop_gamma}).%
}
\label{fig:multiloop_proof1}
\end{supp-figure}
To relate this 
to mfRG, note that the flow of $n$-th order diagrams is only determined
by lower-order diagrams, 
and that the equivalence \ERn{eq:equiv_toshow}
as well as our arguments using generating functions
hold for all orders individually.
Building the series
from the bare interaction,
we can therefore assume the parquet diagrams of the vertex on the r.h.s.\ to be given.
At the one-loop level [\FR{fig:multiloop_flow}(a)],
the definitions for $\dot{\gamma}_{a}^{(1)}$
and $\dot{\mathcal{V}}_1$ are identical,
hence we also have $F_1(n) = P_1(n)$.
For $\dot{\gamma}_{a}^{(2)}$ [\FR{fig:multiloop_flow}(b)], 
the one-loop contribution from the complementary channel,
$\dot{\gamma}_{p}^{(1)}$,
is inserted on the left and right side of the full vertex.
Both of these parts have the same number of diagrams,
which is precisely the number of diagrams 
in $\dot{\mathcal{V}}_2$ (cf.~\FR{fig:multiloop_proof1}).
Hence, we get
$F_2(n) = 2P_2(n)$.
For all higher loops,
$\dot{\gamma}_{a}^{(\ell\textrm{\tiny{+}}2)}$ [\FR{fig:multiloop_flow}(c)],
the previous term is similarly inserted on both sides of the full vertex,
however the center part is constructed
with $\dot{\gamma}_{p}^{(\ell)}$ from loop order $\ell$,
and the proportionality relation becomes more complicated.
We use an inductive argument, starting at $\ell=3$,
and that the number of diagrams
contributing to the lower-loop vertices, 
$\dot{\gamma}_{p}^{(1)}$ and $\dot{\gamma}_{p}^{(2)}$,
is obtained by multiplying the number of diagrams
of the auxiliary vertices by a counting constant
(which keeps track of the different ways to combine vertices at
fixed loop order):
\begin{supp-equation}
F_1(n) = c_1 P_1(n)
\EC \
c_1 = 1
\ESC \quad
F_2(n) = c_2 P_2(n)
\EC \ 
c_2 = 2
\ED
\end{supp-equation}%
Using further the equation illustrated in \FR{fig:multiloop_proof2},
we similarly obtain for all higher loops:
\begin{supp-equation}
F_{\ell+2}(n) = c_{\ell+2} P_{\ell+2}(n)
\EC \quad c_{\ell+2} = 2 c_{\ell+1} + c_{\ell}
\EC
\quad \ell \geq 1
\ED
\end{supp-equation}%
The recursion relation for $c_{\ell}$ with the initial conditions $c_1$ and $c_2$
is known to define the so-called Pell numbers \cite[A000129]{Sloane2017},
which are explicitly given by
\begin{supp-equation}
c_{\ell} = \frac{(1+\sqrt{2})^{\ell} - (1-\sqrt{2})^{\ell}}{2\sqrt{2}}
\ED
\label{eq:Pellnumbers}
\end{supp-equation}%
\begin{supp-figure}[t]
\includegraphics[width=0.48\textwidth]{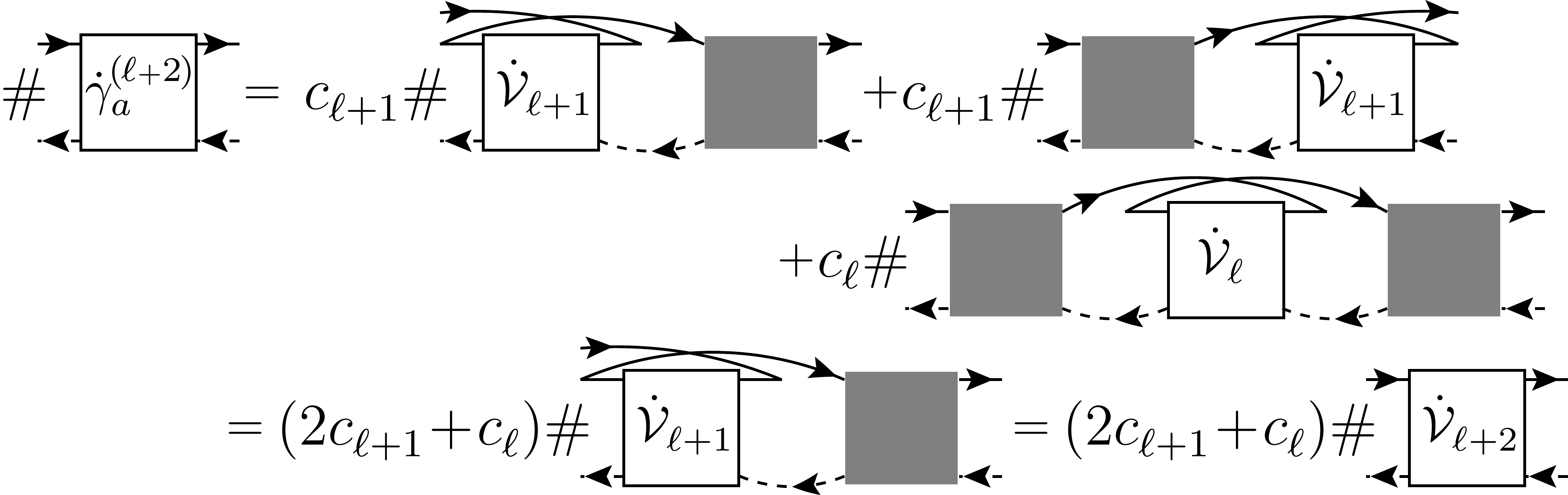}
\caption{%
Relation between the number of diagrams contained in 
$\dot{\gamma}_{a}^{(\ell\textrm{\tiny{+}}2)}$ in $\dot{\mathcal{V}}_{\ell+2}$,
where $\#$ symbolizes that we count the number of diagrams 
of the subsequent vertex.%
}
\label{fig:multiloop_proof2}
\end{supp-figure}
To summarize, the number of diagrams at order $n$ 
of the full vertex,
generated by mfRG 
at loop order $\ell$, is given by
$2 F_{\ell}(n)$,
where $F_{\ell}(n) = c_{\ell} P_{\ell}(n)$,
with generating functions
$f_{\ell}(x) = c_{\ell} p_{\ell}(x)$.
Summing all loops, we find
by using \EsR{eq:aux_genfunc} and \ERn{eq:Pellnumbers}:
\begin{supp-align}
2 \sum_{\ell=1}^{\infty} f_{\ell}(x) & = 
\frac{1}{\sqrt{2}} p_0(x)
\sum_{\sigma = \pm 1} \sigma \sum_{\ell=1}^{\infty} \big[ xp_0(x)(1+\sigma \sqrt{2}) \big]^{\ell} \nonumber \\
& = 
\frac{1}{\sqrt{2}} p_0(x)
\sum_{\sigma = \pm 1} \frac{ \sigma }{1- x p_0(x) (1+\sigma \sqrt{2})} \nonumber \\
& = 
\frac{2xp_0(x)^2}{1-2xp_0(x)-x^2p_0(x)^2}
= q(x)
\EC
\end{supp-align}%
where the last equality follows by repeated use of \ER{eq:defeq}.
Consequently, the sequences corresponding to $q(x)$ and
$2 \sum_{\ell \geq 1} f_{\ell}(x)$ are also equal. 
Using $F_{\ell}(n)=0$ for $\ell \geq n$ [cf.~\ER{eq:aux_genfunc}],
this means
\begin{supp-equation}
P_0(n)(n-1) 
= 2 \sum_{\ell=1}^{\infty} F_{\ell}(n)
= 2 \sum_{\ell=1}^{n-1} F_{\ell}(n)
\ED
\label{eq:SchroederPell}
\end{supp-equation}%
We thus have shown that the number of differentiated diagrams produced by mfRG
at any order $n$
matches the number of differentiated parquet diagrams at this order,
and that an $\ell$-loop fRG flow includes all parquet graphs up to order
$n=\ell+1$.
The details of the proof rely on properties of the XES.
However, generalizing the above strategy to more general models should
be straightforward.

\vspace{1cm}

\section*{S-IV. Relation to paths on a triangular grid}
As a mathematical curiosity, we mention that the
sequences appearing in the previous section have
a certain meaning when counting paths on a triangular grid.
We are not aware of an underlying
connection which goes beyond coincidental properties
of the recursion relations
of the sequences $P_{\ell}(n)$.
Nevertheless, the details are sufficiently intriguing 
that we present them here.
\begin{supp-figure}[t]
\center
\includegraphics[width=.48\textwidth]{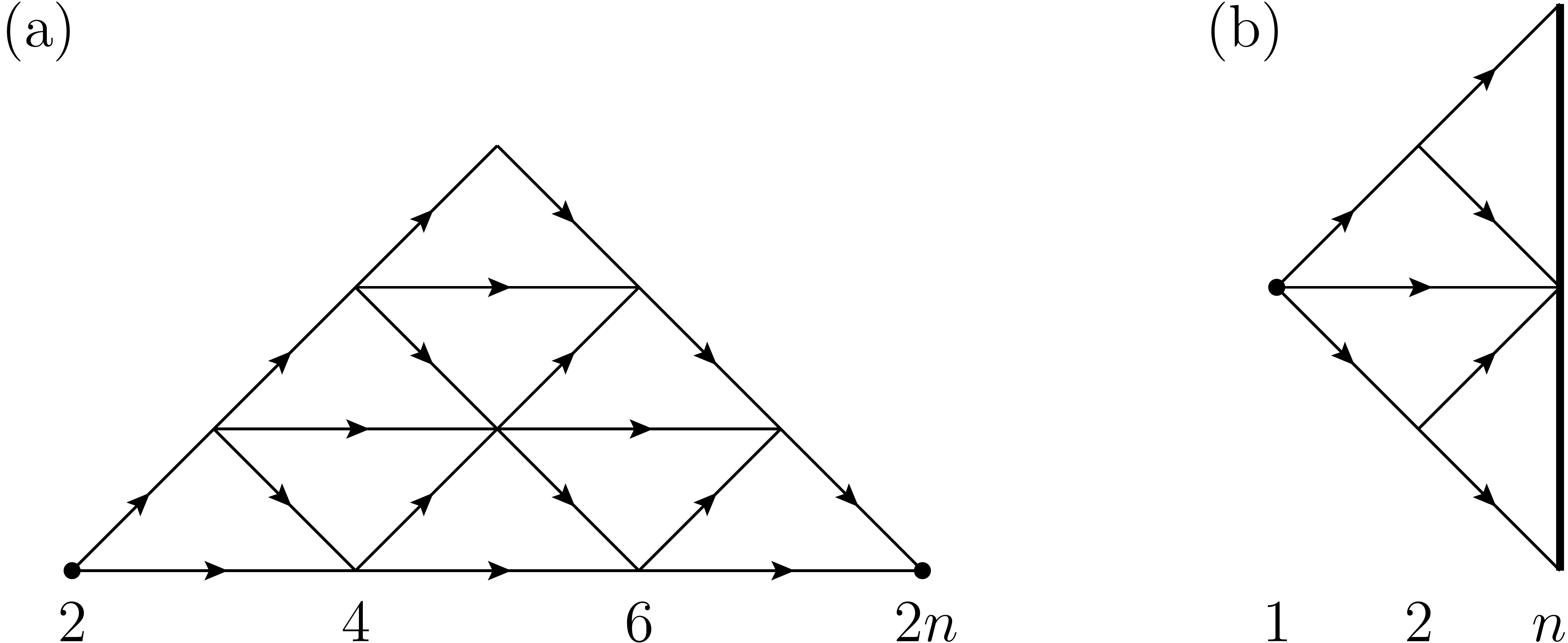}
\caption{%
(a) The (large) Schr\"oder numbers count the number of paths on a triangular grid 
(in the half-plane) between two points on a line. For $n=4$, these are $22$. $16$ of these have a peak at the first level,
$6$ at the second, and only $1$ at the third level [cf.~\ER{eq:SchroederTriangle}].
(b) The Pell numbers count the number of paths on a triangular grid (not restricted to the half-plane)
from a point to a vertical line. For $n=3$, these are $5$.
}
\label{fig:triangular_grid}
\end{supp-figure}
The sequence $P_0(n)$  of \ER{eq:parquet_sequence},
giving the number of parquet graphs
at order $n$, happens to be known in the mathematical literature
by the name
of the (large) Schr\"oder numbers.
These denote the number of paths on a half-triangular grid beginning
and ending on the horizontal axis \cite[A006318]{Sloane2017} [cf.~\FR{fig:triangular_grid}(a)].
The sequences $P_{\ell}(n)$ give the number of these paths
with a peak at level $\ell$ \cite[A006318-A006321]{Sloane2017}, 
or the number of paths starting from the left corner and
ending at level $\ell$ on the right triangle leg (see below).
The Pell numbers [cf.~\ER{eq:Pellnumbers}] count the number of paths on a triangular grid 
(not restricted to a half-plane)
from a point to a vertical line \cite[A000129]{Sloane2017}%
[cf.~\FR{fig:triangular_grid}(b)].
The interpretation for $P_{\ell}(n)$, $\ell \geq 0$,
as paths ending on the right triangle leg
can be understood from a recursion
relation between $P_{\ell}(n)$
with neighboring $\ell$ and $n$ [cf.~\ER{eq:Fln_recursion}].
For this purpose, let us first derive the relation
and construct $P_{\ell}(n)$ as a matrix.
By using \ER{eq:multiloop_sequence} twice and reordering summation indices, we obtain
for $\ell, n \geq 1$:
\begin{supp-align}
P_{\ell+1}&(n+1) = \sum_{m=1}^{n} P_{\ell}(m)P_{0}(n+1-m) \nonumber \\
& = \sum_{m=1}^{n} \sum_{k=1}^{m-1} P_{\ell-1}(k)P_{0}(m-k)P_{0}(n+1-m) \nonumber \\
& = \sum_{m=1}^{n-1} P_{\ell-1}(m) \sum_{k=1}^{n-m} P_{0}(k)P_{0}(n+1-m-k)
\ED
\end{supp-align}%
Via \EsR{eq:parquet_sequence} and \ERn{eq:multiloop_sequence}, this yields
\begin{supp-align}
P_{\ell+1}&(n+1) = \sum_{m=1}^{n-1} P_{\ell-1}(m) [ P_{0}(n+1-m) - P_{0}(n-m)] \nonumber \\
& = \sum_{m=1}^{n-1} P_{\ell-1}(m) P_{0}(n+1-m) - P_{\ell}(n) \nonumber \\
& = \sum_{m=1}^{n} P_{\ell-1}(m) P_{0}(n+1-m) - P_{\ell-1}(n) - P_{\ell}(n) \nonumber \\
& = P_{\ell}(n+1) - P_{\ell-1}(n) - P_{\ell}(n) 
\ED
\end{supp-align}%
We can combine this recursion  
\begin{supp-equation}
P_{\ell}(n+1) = P_{\ell-1}(n) + P_{\ell}(n) + P_{\ell+1}(n+1)
\label{eq:Fln_recursion}
\end{supp-equation}%
with the relation known from \ER{eq:parquet_recursion},
\begin{supp-equation}
P_{0}(n+1) = P_{0}(n) + P_1(n+1)
\EC
\end{supp-equation}%
and \ER{eq:aux_genfunc}, which implies
\begin{supp-equation}
P_{n}(n) = 1
\ESC \qquad
P_{\ell}(n) = 0
\EC \quad \ell \geq n
\ED
\end{supp-equation}%

These equations suffice to build 
the following matrix, defined as $A_{n, \ell} = P_{\ell}(n)$, with $n \geq 1$ 
and $\ell \geq 0$:

\begin{supp-align}
&
\bordermatrix{
					& \!\!\!\!\!\!\!\!\!\!\ell=0,	&	1,			&	2,			&	3,			&	4,			&	5,			&	\dots   \cr
n=1\,\quad			&	1			& 	0			& 	\dots	& 				& 				& 				& 				\cr
\qquad2					&	2			& 	1 			& 	0			& 	\dots	& 				& 				& 			  	\cr
\qquad3					&	6			& 	4			& 	1			&	0			& 	\dots	& 				& 				\cr
\qquad4					&	22		& 	16 		& 	6			& 	1			& 	0		  	& \dots	& 			 	\cr
\qquad5					&	90		& 	68 		& 	30		& 	8			& 	1			& 0			& 	\dots 	\cr
\qquad\,\vdots			 &	\vdots	& 			 	& 				 & 			&				&	\ddots 	&	\ddots	\cr
} \nonumber \\
& \qquad \qquad  \,\,\quad c_{\ell} = 1,\,\quad 2,\,\,\quad 5,\,\quad 12,\,\quad 29
\label{eq:SchroederTriangle}
\end{supp-align}%
If one distorts the matrix slightly, e.g.~by raising
the $\ell$-th column by $\ell$ times half the width between subsequent rows
and ignores all vanishing entries,
one obtains a triangle structure as in \FR{fig:triangular_grid}.
We might consider the entry $A_{0,1}$ as the starting point 
of paths, for which the steps
\begin{supp-align}
& n \to n+1
\EC \qquad
\ell \to \ell
\EC \\ \nonumber
& n \to n+1
\EC \qquad
\ell \to \ell+1
\EC \\ \nonumber
& n \to n
\EC \qquad
\ell+1 \to \ell
\end{supp-align}%
are allowed.
Then, the entry $A_{n,\ell}$ indeed
gives the number of such paths ending at the corresponding point
on the triangular grid.
The equality between the number of differentiated parquet
and mfRG diagrams shown in Sec.~S-III, \ER{eq:SchroederPell}, translates into
\begin{supp-equation}
(n-1) A_{n, 0} = 2 \sum_{\ell=1}^{n-1} c_{\ell} A_{n, \ell}
\ED
\label{eq:SchroederPell_triangle}
\end{supp-equation}%
While many relations for the matrix $A$ [\ER{eq:SchroederTriangle}] 
are known \cite[A033877]{Sloane2017},
we have not found a proof of \ER{eq:SchroederPell_triangle}
in the literature. 

\end{document}